\documentclass[12pt]{article}
\usepackage{amsmath}
\usepackage{amssymb}
\usepackage{amsthm}
\usepackage{amsfonts}
\usepackage{algorithm}
\usepackage{algorithmic}
\usepackage{graphicx}
\usepackage{pdfpages}
\usepackage{multirow}

\usepackage{xcolor,colortbl}
\usepackage{multicol}
\usepackage{url}
\usepackage{subfigure}
\usepackage{listings}
\definecolor{Gray}{gray}{0.92}
\definecolor{LightCyan}{rgb}{0.88,1,1}

\newcolumntype{a}{>{\columncolor{Gray}}c}
\newcolumntype{b}{>{\columncolor{white}}c}

\usepackage{bbm,epsfig,graphics,epic,color,rotating,color}

    \usepackage[left=2.2cm,right=2.2cm,top=2cm,bottom=2cm]{geometry}

\numberwithin{equation}{section}

\newtheorem*{theorem*}{Theorem}

\theoremstyle{definition}

\title{Lasso regularization for mixture experiments with noise variables.}
\author{Manuel Gonz\'alez-Navarrete\thanks{Departamento de Matem\'atica y Estad\'i{}stica, Universidad de La Frontera. Avda. Francisco Salazar 01145, Temuco, Chile. E-mail address: manuel.gonzaleznavarrete@ufrontera.cl}, \\ Fabi\'an Manr\'iquez-M\'endez\thanks{Instituto de Estad\'istica, Universidad de Valpara\'iso. Av. Gran Bretaña 1111, Valpara\'iso, Chile. E-mail address: fabian.manriquez@postgrado.uv.cl} \ and Manuel Pereira-Barahona\thanks{Departamento de Estad\'i{}stica, Universidad del B\'io-B\'io. Avda. Collao 1202, Concepci\'on, Chile. E-mail address: mpereira@ubiobio.cl}}

\date{}

\begin{document}

\maketitle

\thispagestyle{empty} %
\vspace{2pt}

\begin{abstract}
We apply classical and Bayesian lasso regularizations to a family of models with the presence of mixture and process variables. We analyse the performance of these estimates with respect to ordinary least squares estimators by a simulation study and a real data application. Our results demonstrate the superior performance of Bayesian lasso, particularly via coordinate ascent variational inference, in terms of variable selection accuracy and response optimization.
\end{abstract}

\bigskip


%

\section{Introduction}\label{sec1}

The exploration of complex systems through mixture experiments and the influence of external process variables represents a significant challenge in various fields of science and engineering. The ability to predict and optimize responses in such systems is crucial for technological advancement and innovation \cite{EXP,EXP2}.

In a mathematical framework, the study of mixture experiments involves building a model describing the relationship among the response and the mixture and process variables. This task requires the choose of an experimental design, and the fit of the statistical model by employing the data collected after experimentation. The usual tools to estimate model parameters are the ordinary least squares \cite{Azc,EXP} and, to a lesser extent, the partial least squares \cite{Kett1992,Mutt2007}.

In this context, regularization techniques such as lasso and its Bayesian extension stand out as fundamental tools for model analysis and selection in statistical literature \cite{jamesetal}, being good candidates for challenges such as high-dimensional mixture experiments. Lasso, introduced by \cite{Tib}, marked an advancement in regression by proposing a technique that minimizes the sum of squared residuals with a constraint on the $L^1$ norm of the coefficients, facilitating variable selection and reducing model complexity. Bayesian lasso, proposed by \cite{park2008bayesian}, extends this approach by incorporating a Bayesian perspective that assigns Laplace prior distributions to the regression parameters. This innovation maintains the effectiveness of lasso in variable selection while introducing Bayesian flexibility in estimation. Subsequent developments, such as those by \cite{BH}, \cite{Bmallick}, and \cite{bAlhamzawi}, have delved into the hierarchical structure of the model and improved inference algorithms, highlighting the robustness of Bayesian lasso in variable selection and predictive accuracy.

This study focuses on the integration of mixture experiments with process variables and the application of lasso regularization techniques to simultaneously optimize the mean and variance of the response. In particular, we explore the mixture-process models with noise variables described in \cite{EXP}, emphasizing their importance in understanding the interaction between mixture components and process conditions. The mathematical formulation underlying the optimization of these models is discussed, aiming to evaluate the performance of classical formulation and Bayesian lasso by employing Markov chain Monte Carlo \cite{turkman2019computational} and variational approximation methods \cite{blei2017variational}. In this sense, a practical application of these concepts is illustrated through a simulation study to evaluate their performances in variable selection task. Moreover, a real data example from \cite{GBM} is included, discussing the effectiveness of the proposed approach in the study of mixture experiments.

The rest of the paper is organized as follows. In Section \ref{Sec:thba} we introduce the theoretical aspects in the mathematical study of mixture experiments and the proposed regularization methods. Section \ref{Sec:sim} includes the results of a simulation study to evaluate the performance of such methods. In Section \ref{Sec:app} we expose an application for a soap production experiment. Finally, Section \ref{Sec:con} contains some conclusions.

\section{Theoretical Background}
\label{Sec:thba}

\subsection{Mixture experiments with noise variables}

Mixture models with process variables represent an advanced tool for analyzing systems where both component proportions and specific external conditions (process variables) influence the system's response. These models extend traditional frameworks by incorporating additional variables that reflect the conditions under which the experiment is conducted.

The general formulation of a model including mixture components $\mathbf{x}$, process variables $\mathbf{w}$, and possibly noise variables $\mathbf{z}$, is described by the equation:
\begin{equation}
\label{generalmodel}
	\begin{split}
		Y  = &f( \textbf{x},\textbf{w}, \textbf{z}) = \displaystyle\sum_i \alpha_i x_i + \mathop{\sum\sum}_{i<j}  \alpha_{ij} x_i x_j + \displaystyle\sum_i\sum_p  \delta_{ip} x_i w_p \\
		& + \mathop{\sum\sum}_{i<j}\sum_p \delta_{ijp} x_i x_j w_p + \displaystyle\sum_i\sum_t  \gamma_{it} x_i z_t + \mathop{\sum\sum}_{i<j}\sum_t \gamma_{ijt} x_i x_j z_t \\
		& + \displaystyle\sum_i\sum_p\sum_t  \eta_{ipt} x_i w_p z_t + \mathop{\sum\sum}_{i<j}\sum_p\sum_t \eta_{ijpt} x_i x_j w_p z_t + \varepsilon
	\end{split}
\end{equation}
where $Y$ is the response variable; $\underline{\beta}=(\underline{\alpha}, \underline{\delta}, \underline{\gamma}, \underline{\eta})$ is the vector of coefficients modeling linear effects and interactions; and $\varepsilon$ represents the error term, normally distributed with mean zero and variance $\sigma^2$.

Key constraints for such models include \cite{GBM}:
\begin{itemize}
	\item The proportions of mixture components $\mathbf{x}$ must sum to 1, i.e., $\sum_i x_i = 1$, ensuring that the model adequately reflects the nature of mixtures.
	\item Mixture components $x_i$ and process variables $w_p$ should be selected to faithfully reflect the system under study, including only those factors that have a significant impact on the response.
	\item Conveniently, the noise variables are supposed to be independent and identically distributed, with $\mathbb{E}(Z_t) = 0$ and $\mathbb{V}(Z_t) =1$.
\end{itemize}


The usual tool to estimate parameters in \eqref{generalmodel} is the method of ordinary least squares, which are given by
\begin{equation}
    \hat{\underline{\beta}}_{OLS}= {\arg \min}_{\underline{\beta}}(y-X\underline{\beta})^t(y-X\underline{\beta})
\end{equation}
Once we obtain the fitted model for response variable $Y$, the objective is finding optimal configurations of mixture variables $\mathbf{x}$ and process and noise variables $\mathbf{w}$ and $\mathbf{z}$, respectively, which optimize the response for the experiments. This task is completed by using the desirability function approach proposed by Derringer and Suich \cite{DS}, and extensively used in the recent literature \cite{Azc,des1,des2,des3}.

The desirability function is defined for estimated response functions, such as the moments of $Y$, $\mathbb{E}(Y^n)$. The values of these functions increase as the "desirability" of the corresponding response increases. In this sense, for instance, the desirability function of the expectation of a function of the response variable is given by
\begin{equation}
\label{di}
d \left(\widehat{\mathbb{E}}(g(Y))\right)=
\begin{cases}
0, & \text{if} \ \widehat{\mathbb{E}}(g(Y)) \le \mathbb{E}(g(Y))_*\\
\left[\frac{\widehat{\mathbb{E}}(g(Y))- \mathbb{E}(g(Y))_*}{\mathbb{E}(g(Y))^*-\mathbb{E}(g(Y))_*}\right]^r, & \text{if} \ \mathbb{E}(g(Y))_*< \widehat{\mathbb{E}}(g(Y)) < \mathbb{E}(g(Y))^*\\
1, & \text{if} \ \widehat{\mathbb{E}}(g(Y)) \ge \mathbb{E}(g(Y))^*
\end{cases}
\end{equation} 
 
The values $\mathbb{E}(g(Y))_*$ and $ \mathbb{E}(g(Y))^*$ give the minimum and maximum acceptable values of $\mathbb{E}(g(Y))$, respectively. The parameter $r$ is arbitrarily chosen. Finally, the individual desirabilities are combined using the geometric mean,
 \begin{equation}
 \label{D}
     D(\textbf{x},\textbf{w}, \textbf{z}) = \left(d\left(\widehat{\mathbb{E}}(g_1(Y))\right) \cdot \ldots \cdot  d\left(\widehat{\mathbb{E}}(g_d(Y))\right)\right)^{1/d}
 \end{equation}
This single value of $D$ is maximized to obtain the overall assessment of the desirability of the combined expected response functions. In particular, we use $g_1(Y) = Y$ and $g_2(Y) = -(Y -\mathbb{E}(Y))^2$. In other words, we maximize the expectation and minimize the variance of the response variable $Y$.

  \subsection{Lasso regularization}
The introduction of lasso, as proposed by Tibshirani (1996), represented a significant step forward in regression techniques. By minimizing the sum of squared residuals under a constraint on the $L^1$ norm of coefficients, lasso facilitates variable selection and effectively reduces model complexity. Building upon this foundation, Bayesian lasso, as introduced by Park and Casella (2008), takes regression analysis further by adopting a Bayesian framework that assigns Laplace prior distributions to regression parameters. This Bayesian perspective not only preserves the variable selection capabilities of lasso but also introduces greater flexibility in parameter estimation.


  \subsubsection{Classical formulation}

\par The lasso is a form of penalized least squares that minimizes the residual sum of squares while controlling the $L^1$ norm of the coefficient vector $\underline{\beta}$. The lasso estimator for a classical regression model is given by,
\begin{equation}
    \hat{\underline{\beta}}_L= {\arg \min}_{\underline{\beta}}(y-X\underline{\beta})^t(y-X\underline{\beta})+ \lambda ||\underline{\beta}||_1
\end{equation}
where $\lambda \geq 0$ is called the shrinkage parameter. In the case $\lambda=0$, we have $\hat{\underline{\beta}}_{L}=\hat{\underline{\beta}}_{OLS}$, the ordinary least squares (OLS) estimation, and sufficiently large $\lambda$ reduces $\underline{\beta}_{L}$ to zero. The lasso has a Bayesian interpretation \cite{Tib}, since the lasso estimation can be seen as the mode of the posterior distribution of $\underline{\beta}$, when double-exponential and independent prior distributions are assigned to the $p$ regression coefficients,
\begin{equation}
    p(\underline{\beta} \mid  \tau)= (\tau /2)^p \exp \left ( -\tau ||\underline{\beta}||_1  \right )
\end{equation}
where $p(\underline{y} \mid \underline{\beta},\sigma^2)=\mathcal{N}(\underline{y} \mid X \underline{\beta},\sigma^2 \textbf{I}_n)$, for any fixed values of $\sigma>0$ and $\tau>0$, with penalty $\lambda=2 \tau \sigma^2$.

\subsubsection{Bayesian formulation}
The work \cite{park2008bayesian} shows a Bayesian formulation of lasso regression. The hierarchical model is defined by:
\begin{equation}
\label{eq:lasso}
\begin{split}
        \underline{y} \mid X , \underline{\beta} ,\sigma^{2} & \sim \mathcal{N}(X \cdot \underline{\beta},\sigma^{2} \cdot I_n) \\
      \underline{\beta} \mid \sigma^{2}, \underline{\tau}  & \sim \mathcal{N}(\underline{0},\sigma^{2} \mathbf{D}_{\tau}) \\
      \tau_j \mid \lambda & \sim \operatorname{Exp}(\lambda) \quad j=1,\ldots,p
\end{split}
\end{equation}
where $\mathbf{D}_{\tau}=\operatorname{diag}(\tau_1,\ldots,\tau_p)$ y $\tau_j \mid \lambda$  and $\tau_j$ are conditionally independent for all $j$. The model can be completed with the gamma prior distributions $(\sigma^{2})^{-1} \sim Ga(a_0,b_0)$ and $\lambda \sim Ga(c_0,d_0)$, where $a_0, b_0, c_0$ and $d_0$ the hyperparameters. Let $\underline{\theta}=\left (  \underline{\beta},\sigma^{2},\underline{\tau},\lambda \right )$ be the vector of the parameters for this model. The posterior distribution will be proportional to the model distribution times the prior distribution for the latent components and the parameters:
\begin{equation*}
    p(\underline{\theta} \mid \underline{y},X) \propto p(\underline{y} \mid X,\underline{\beta},\sigma^{2}) p(\underline{\beta} \mid \sigma^{2},\underline{\tau}) p(\underline{\tau} \mid \lambda) p(\sigma^{2}) p(\lambda)
\end{equation*}
The posterior distribution for hierarchical model is often intractable.

\subsection{Bayesian model estimation}

In this section we explain the point estimator we use, the variable selection methods and the computational tools to approximate the posterior distributions. In particular, we include the Markov chain Monte Carlo (MCMC) and variational alternatives by coordinate ascent variational inference (CAVI) and automatic differentiation variational inference (ADVI).

\subsubsection{Point estimation and variable selection in Bayesian lasso}


We adopt the use of posterior mean, $\hat{\theta} = \mathbb{E}(\underline{\theta} \mid \underline{y},X)$, to give point estimations. The choice of this estimator is driven by its capacity to condense all the information provided by the posterior distribution, offering an estimate that considers the diversity of possible parameter values, in contrast to the one-dimensional approach of the MAP \cite{park2008bayesian}.

Furthermore, Bayesian lasso provides interval estimates that can guide variable selection. Usually, for each parameter, it is used a 95\% credible interval and if the interval contains the value zero, then the regression coefficient is excluded \cite{park2008bayesian}. This criterion will be denoted by CI (credible interval).

However, as discussed by \cite{BNet}, the 95\% credible intervals are usually too wide and most predictors would consequently be removed. Therefore, we use the criterion proposed by Li and Lin, that is, we consider the posterior probability of the interval $\left[-\sqrt{\mathbb{V}(\beta_j) }| \underline{y}, X ; \sqrt{\mathbb{V}(\beta_j) }| \underline{y}, X \right]$. In this sense, a regression coefficient is excluded if such probability exceeds a certain threshold and is retained otherwise. In particular, we use $0.5$ as a threshold. This criterion is called scaled neighborhood (SN).

\subsubsection{Markov chain Monte Carlo}

In the field of Bayesian inference, Markov chain Monte Carlo (MCMC) is the most common method to approximate posterior distributions \cite{turkman2019computational}. However, usual MCMC includes high autocorrelation and convergence is slow especially in high-dimensional spaces. To overcome this problem, the rstan package, implemented in R (see \cite{stan}), uses an advanced version of Hamiltonian Monte Carlo (HMC) known as the no-U-turn sampler (NUTS) (see \cite{hoffman14amcmc}), optimizing the sampling process by eliminating the need to manually adjust the number of steps in each improving efficiency in parameter space exploration, reducing autocorrelation between samples and speeding up convergence.

NUTS is an extension of the HMC algorithm that solves the problem of selecting the optimal number of simulation steps. It employs a recursive strategy that automatically expands the trajectory in parameter space, stopping when a reversal or "U-turn" is detected in the trajectory, hence its name. This enhances sampling efficiency by reducing autocorrelation between samples and optimizing the use of computational resources.

Formally, updating the parameters \(\theta\) in NUTS can be described using Hamiltonian dynamics, where an auxiliary momentum \(p\) is introduced, and the system's evolution is simulated under the Hamiltonian \( H(\theta, p) = U(\theta) + K(p) \). Here, \( U(\theta) \) represents the negative logarithmic potential of the posterior distribution, and \( K(p) \) is the kinetic energy associated with the momentum \( p \), typically defined as \( K(p) = \frac{1}{2} p^T M^{-1} p \), where \( M \) is the mass (or covariance) matrix that can be adjusted to reflect parameter scales (see \cite{hoffman14amcmc}).
\begin{equation}
	\theta_{n+1}, p_{n+1} = \text{Leapfrog}(\theta_n, p_n, \epsilon_n, L_n)
\end{equation}
where \(\text{Leapfrog}(\cdot)\) denotes the leapfrog integration steps used for numerical simulation of Hamiltonian dynamics, \( \epsilon_n \) is the adaptive step size, and \( L_n \) is the number of leapfrog steps, determined dynamically \cite{bishop2006pattern}.

The implementation of NUTS in rstan allows for more efficient Bayesian inference in high-dimensional models and reduces manual intervention in selecting sampler hyperparameters. The MCMC algorithms, specifically NUTS, were run until convergence, evaluated using the \(\hat{R}\) statistic (potential scale reduction), ensuring \(\hat{R} < 1.1\) (see \cite{rhat}).

\subsubsection{Methods based on variational inference}

The concept behind variational inference methods is to propose a family of densities and find a member $q$ of that family which closely approximates the target posterior $p(\underline{\theta} \mid \underline{y},X)$ \cite{blei2017variational}. In other words, instead of computing the true posterior, we endeavor to determine the parameters $\phi$ of a particular distribution $q^*$ (the approximation to our true posterior) such that
\begin{equation}
\label{eq:argmin}
    q^*= \arg \min \mathcal{L}(q(\underline{\theta} ; \phi) 
\ || \ p(\underline{\theta} \mid \underline{y},X))   
\end{equation}
where $\mathcal{L}(\cdot \ || \ \cdot)$ denote the Kullback-Leibler divergence, given by
 $\mathcal{L}(q(\underline{\theta};\phi) \ || \ p(\underline{\theta} \mid \underline{y},X))= \mathbb{E}_{q} \left [ \ln \frac{q(\underline{\theta} ;\phi)}{p(\underline{\theta} \mid \underline{y},X)} \right ]$. Therefore \eqref{eq:argmin} is equivalent to maximizing
\begin{equation}
\label{eq:vari_q}
    q^{*}= \arg \max \{  \underbrace{\mathbb{E}_q \left [ \ln p(\underline{y} , \underline{\theta},X) -\ln q(\underline{\theta};\phi) \right ]}_{ELBO} \}
\end{equation}
This expression is called evidence lower bound (ELBO).
In particular, we want to optimize the ELBO in mean field variational inference, that is, the joint distribution reduces to the product of marginal distributions, $q( \underline{\theta}) = \prod_{i=1}^p q(\theta_i)$.
\begin{itemize}
    \item Coordinate ascent variational inference: This algorithm to solve the optimization problem was introduced by
\cite{bishop2006pattern} and denoted by coordinate ascent variational inference (CAVI). The CAVI optimizes one factor of the mean field variational density at a time. This is defined as an iterative optimization of $q_j$ for $j=1,\ldots,p$, while the other variational distributions are fixed. The optimal $q_j$ is proportional to the exponential of the log of the complete conditional distribution, is given by
\begin{equation}
\label{}
  q(\theta_j) \propto \exp  \{   \mathbb{E}_{\theta_{-(j)}}  [ \ln p(\theta_j \mid \theta_{-(j)},\underline{y},X)  ] \} \quad j =1,\ldots,p
\end{equation}
where $ \theta_{-(j)}=(\theta_1,\ldots,\theta_{j-1}.\theta_{j+1},\ldots,\theta_p)$.

In the context of CAVI, the focus lies on iteratively adjusting the parameters within the variational distribution until certain convergence standards are reached. This process entails performing analytic derivations for the updates, which may prove to be time-intensive at most and impractical in certain scenarios. The main objective is to optimize the ELBO in the mean field variational inference.

For the Bayesian lasso model proposed in \eqref{eq:lasso}, the variational posterior for $\underline{\beta}$ and $\sigma^{2}$, is given by (for details, see \cite{alves2021variational})
\begin{equation}
  \notag   \begin{split}
     q(\underline{\beta},\sigma^{2}) & =\mathcal{N}(\underline{\beta} \mid m_\beta,\sigma^{2}\cdot C_\beta) Ga ((\sigma^{2})^{-1} \mid a_0,b_0)
    \end{split}
\end{equation}
it is recognized that is a normal-gamma distribution with parameters:
\begin{equation*}
        C^{-1}_{\beta} = \mathbb{E} \left[ D_{\tau}^{-1}  \right] + X^t X, \quad  m_{\beta}=C_\beta X^t y,  \quad a_{\sigma^{2}}=a_0+\frac{1}{2} \quad  \text{and} \quad  b_{\sigma^{2}} =b_0+\frac{1}{2} \left ( y^t y-m_{\beta}^t C_{\beta}^{-1}m_{\beta} \right ).
\end{equation*}
the variational distribution for $\tau_j$, is given by $q(\tau_j)= \mathcal{GIG}(\tau_j \mid c_\tau,d_\tau,f_{\tau_j})$, where $\mathcal{GIG}$ is a generalized inverse Gaussian distribution, with parameters $c_\tau = \frac{1}{2}$,  $d_\tau = 2 \mathbb{E}_\lambda \left [ \lambda \right ]$ and $f_{\tau_j}=\mathbb{E}_{\sigma^{2} \beta}\left [ (\sigma^{2})^{-1} \beta_j^2 \right ].$
Finally, the variational distribution for $\lambda$, is given by
\begin{equation}
 \notag    q(\lambda)=Ga( \lambda \mid a_\lambda,b_\lambda )
\end{equation} 
it is recognized that is a gamma distribution with parameters
\begin{equation}
 \notag
a_\lambda=  g_0+p  \quad \text{and} \quad b_\lambda= h_0 + \sum_{j=1}^{p} \mathbb{E}\left [ \tau_j \right ].
\end{equation}

\item  Automatic differentiation variational inference (ADVI):
Implementing CAVI requires careful thought about the target distribution and choosing an appropriate variational family specific to the problem. Alternatively, \cite{ADVI} offer a way to automate variational inference. We will first assume all model parameters are continuous. In ADVI, the ELBO is first re-written as
\begin{equation*}
    \operatorname{ELBO}(\underline{y},\phi):= \mathbb{E}_q  \left [\ln p(\underline{y},T^{-1}(\zeta),X)+\ln | J_{T^{-1}}(\zeta)|- \ln q(\zeta;\phi)  \right ]
\end{equation*}
Here, $T$ is a function that transforms $\theta$ to $\zeta$, where $\zeta \in \mathbb{R}^{dim(\theta)}$. That is, $T:\operatorname{support}(\theta) \rightarrow \mathbb{R}^{dim(\theta)}$, identified as $\zeta=T(\theta)$ and $J_{T^{-1}}(\zeta)$ is the
Jacobian of the inverse of T. As all the model parameters 
$\zeta$ have support on the real line, a suitable variational distribution for $\zeta$ is a normal distribution. Using a multivariate Gaussian variational distribution $q(\zeta; \phi ) = N(\zeta | m, LL^t)$ is specified for $\zeta$ and the variational parameters are $\phi = (m, L)$ enables us to compute the expectation and its gradient using a Monte Carlo estimate.
Specifically, to estimate the ELBO, one can sample values from the variational distributions and evaluate the expression inside the expectation mentioned above.
To maximize the ELBO, the gradient of the ELBO with respect to the variational parameters is required. That is
\begin{equation*}
  \nabla_\phi  \operatorname{ELBO}(\underline{y},\phi):= \nabla_\phi \mathbb{E}_q  \left [\ln p(\underline{y},T^{-1}(\zeta),X)+\ln | J_{T^{-1}}(\zeta)|- \ln q(\zeta;\phi)  \right ]
\end{equation*}
Once again, we can assess this through Monte Carlo integration. However, computing the gradient of a random variate isn't straightforward. Hence, it's prudent to initially draw a standard normal random variable and then scale it by the variational standard deviation and mean. This way, we can incorporate the gradient within the expectation. To clarify further:

\begin{equation}
\label{eq:aprox_elbo}
  \nabla_\phi  \operatorname{ELBO}(\underline{y},\phi) \approx \frac{1}{S} \sum_{s=1}^{S}  \nabla_\phi \left [\ln p(\underline{y},T^{-1}(\zeta),X)+\ln | J_{T^{-1}}(\zeta)|- \ln q(\zeta;\phi)\textbar_{\zeta=m+L \epsilon^{(s)}}
  \right ]
\end{equation}
where $\epsilon^{(s)} \sim N(0,I), s=1, \ldots,S.$ One can also easily compute the stochastic
gradient approximation of \eqref{eq:aprox_elbo} 
\end{itemize}
For the variational inference methods, convergence was determined by monitoring the ELBO. Specifically, convergence was achieved when the relative change in the ELBO between successive iterations fell below a predefined tolerance threshold. This criterion ensured that the optimization process had sufficiently stabilized, indicating that the variational approximation was close to the true posterior distribution (see \cite{ADVI}).

\section{A simulation study for variable selection}
\label{Sec:sim}

In this section we analyse the performance of lasso regularization in the variable selection task. We include the results for classical lasso and Bayesian lasso methods (CAVI, ADVI and MCMC), with variable selection criteria CI and SN.

We suppose the experiments are governed by a reduced form of the quadratic mixture model, given by \eqref{generalmodel}, with $i=1, 2, 3$,  $p=1$ and $t=2$. We consider a model where $\underline{\beta} = ( \underline{\alpha}, \underline{\delta}, \underline{\eta})$, for which we set $\underline{\alpha}= \underline{\delta} = \underline{1}$ and $\underline{\eta} = \underline{0}$, and then evaluate the performance of lasso to set $\underline{\eta}$ equal zero.

\subsection{Data Generation}
The primary predictors \( x_1, x_2, \) and \( x_3 \) were generated under constraints to ensure their sum is 1. Specifically, \( x_1 \) and \( x_2 \) were drawn from uniform distributions \( U(0.2, 0.8) \) and \( U(0.15, 0.5) \) respectively, while \( x_3 \) was determined as \( 1 - x_1 - x_2 \), ensuring it lies within the range \( [0.05, 0.3] \).

Additional predictors \( w_1, z_1, \) and \( z_2 \) were introduced to simulate the effects of process and noise variables. The variable \( w_1 \) was sampled from a binary distribution taking values \( 0.5 \) and \( 1 \) with equal probability. Both \( z_1 \) and \( z_2 \) were drawn from standard normal distributions.

The response variable \( Y \) was then generated based on the reduced model, with an added noise term \( \varepsilon \) drawn from a normal distribution with mean 0 and standard deviation $\sigma=0.5$.

The implementation of the proposed Bayesian methods requires careful hyperparameter selection and convergence criteria. In this work, we used a prior distribution configuration that includes gamma distributions for \(\phi\) and \(\lambda\), and exponential distributions for \(\tau\). Initial values for the Markov chains were based on previous estimates obtained via ordinary least squares (OLS) to improve sampling efficiency.  These criteria aim to ensure robust and accurate estimation of the model parameters.

\subsection{Results}

For each method, the frequency of variable selection across the 1000 simulations was recorded. Table \ref{tab:freq_simulation} shows detailed results on the variable selection for the simulation study. In our simulation, methods with larger $N(\alpha)$ and $N(\delta)$ and smaller $N(\eta)$ are considered to perform better.

\begin{figure}
\begin{center}
		\includegraphics[width=1\textwidth]{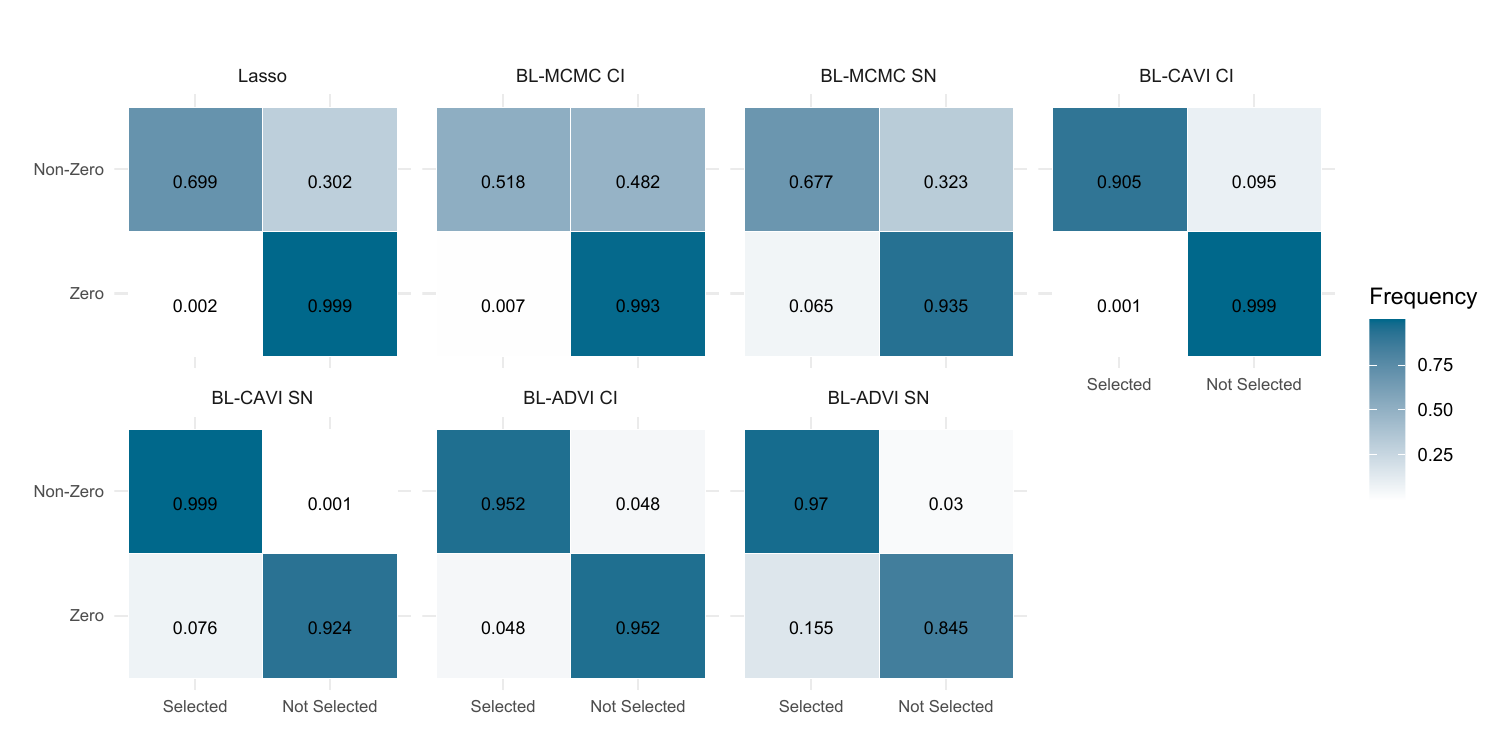} 
\end{center}
	\caption{The confusion matrices for lasso regularization methods.}
	\label{confusion}
\end{figure}

The confusion matrices in Figure \ref{confusion} underscore that CAVI outperforms other methods with the highest true positives and lowest false negatives, indicating superior parameter selection accuracy. Conversely, MCMC variants show notably poorer performance, highlighting their inefficacy in accurate parameter identification.

\begin{table}[ht]
\centering
\begin{tabular}{| c | c | c | c | c | c | c | c | }
\hline
 & \multirow{2}{*}{LASSO} & \multicolumn{2}{c|}{BL-MCMC} & \multicolumn{2}{c|}{BL-CAVI} & \multicolumn{2}{c|}{BL-ADVI} \\[0.1cm]
 \cline{3-8}
   & &   CI & SN & CI & SN & CI & SN \\
\hline
$N(\alpha_{1})$ & 1.000  & 1.000 & 1.000 & 1.000 & 1.000 & 0.998 & 0.998 \\ 
$N(\alpha_{2})$ & 1.000  & 0.905 & 0.995 & 1.000 & 1.000 & 0.998 & 0.998 \\ 
$N(\alpha_{3})$ & 1.000  & 0.640 & 0.963 & 1.000 & 1.000 & 0.998 & 0.998 \\ 
$N(\alpha_{12})$ & 1.000  & 0.527 & 0.850 & 1.000 & 1.000 & 0.991 & 0.995 \\  
$N(\alpha_{23})$ & 0.997  & 0.007 & 0.039 & 0.856 & 0.999 & 0.857 & 0.913 \\  
$N(\alpha_{13})$ & 0.023  & 0.507 & 0.633 & 0.958 & 1.000 & 0.947 & 0.960 \\ 
$N(\delta_{11})$ & 1.000  & 1.000 & 1.000 & 1.000 & 1.000 & 0.998 & 0.998 \\ 
$N(\delta_{21})$ & 1.000  & 0.699 & 0.964 & 1.000 & 1.000 & 0.998 & 0.998 \\  
$N(\delta_{31})$ & 1.000  & 0.438 & 0.949 & 0.955 & 1.000 & 0.994 & 0.997 \\ 
$N(\delta_{121})$ & 0.051  & 0.489 & 0.607 & 0.992 & 1.000 & 0.977 & 0.986 \\   
$N(\delta_{231})$ & 0.096  & 0.003 & 0.028 & 0.394 & 0.986 & 0.763 & 0.852 \\   
$N(\delta_{131})$ & 0.215  & 0.006 & 0.099 & 0.709 & 0.999 & 0.901 & 0.941 \\ 
$N(\eta_{111})$ & 0.002  & 0.017 & 0.121 & 0.000 & 0.101 & 0.100 & 0.283 \\  
$N(\eta_{211})$ & 0.001  & 0.004 & 0.022 & 0.000 &  0.084 & 0.092 & 0.210 \\ 
$N(\eta_{311})$ & 0.000  & 0.004 & 0.010 & 0.000 & 0.040 & 0.047 & 0.158 \\  
$N(\eta_{112})$ & 0.003  & 0.016 & 0.203 & 0.003 & 0.122 & 0.095 & 0.268 \\ 
$N(\eta_{212})$ & 0.001  & 0.006 & 0.094 & 0.002 & 0.101 & 0.075 & 0.216 \\  
$N(\eta_{312})$ & 0.000  & 0.007 & 0.047 & 0.000 & 0.052 & 0.051 & 0.157 \\   
$N(\eta_{1211})$ & 0.002  & 0.005 & 0.036 & 0.004 & 0.092 & 0.035 & 0.132 \\  
$N(\eta_{2111})$ & 0.001  & 0.002 & 0.015 & 0.000 & 0.054 & 0.000 & 0.063 \\  
$N(\eta_{1311})$ & 0.002  & 0.003 & 0.013 & 0.001 & 0.042 & 0.018 & 0.084 \\  
$N(\eta_{1212})$ & 0.001  & 0.007 & 0.098 & 0.002 & 0.110 & 0.031 & 0.124 \\  
$N(\eta_{2112})$ & 0.003  & 0.005 & 0.062 & 0.000 & 0.048 & 0.006 & 0.059 \\ 
$N(\eta_{1312})$ & 0.002  & 0.007 & 0.059 & 0.000 & 0.064 & 0.022 & 0.101 \\  
   \hline
\end{tabular}
   \caption{Frequency of retaining for the regression coefficients.}
    \label{tab:freq_simulation}
\end{table}

Complementary, to evaluate the effectiveness of lasso and Bayesian lasso methods in the context of simulations, we propose the use of the \textit{balanced accuracy index (BAI)}, introduced by \cite{F1}. The \textit{BAI} improves evaluation in contexts where variable selection proportions may be unbalanced, it is calculated as the average of the true positive rate and the true negative rate. Specifically, it is defined as:
\begin{equation*}
    BAI = \frac{1}{2}\left(\frac{TP}{TP + FN} + \frac{TN}{TN + FP}\right),
\end{equation*}
where $TP$ is the number of truly non-zero parameters correctly selected, $FP$ indicates the number of truly zero parameters incorrectly selected, $FN$ represents the number of truly non-zero parameters incorrectly excluded, and $TN$ is the number of truly zero parameters correctly not selected.

Table \ref{tab:freq_simulationF1} shows that, according to the \textit{BAI}, CAVI-SN stands out as the most efficient method for variable selection in the context of simulations, closely followed by CAVI and ADVI with CI criterion, which also demonstrate high efficiency. The classical lasso and ADVI-SN methods show good performance, whereas MCMC methods, perform less effectively. These results suggest that Bayesian variants of lasso, particularly CAVI, may offer significant advantages in terms of variable selection performance, especially in datasets where the balance between sensitivity and specificity is important.

\begin{table}[ht]
    \centering
\begin{tabular}{| c | c | c | c | c | c | c | c | }
\hline
 & \multirow{2}{*}{LASSO} & \multicolumn{2}{c|}{BL-MCMC} & \multicolumn{2}{c|}{BL-CAVI}& \multicolumn{2}{c|}{BL-ADVI} \\[0.1cm]
 \cline{3-8}
   & &   CI & SN & CI & SN & CI & SN \\
\hline
\textit{BAI} & 0,849 & 0,756 & 0,806 & 0,952 & 0,961 & 0,952 & 0,907 \\
	\hline
\end{tabular}
    \caption{Comparison of methods performance by using the balanced accuracy index.}
    \label{tab:freq_simulationF1}
\end{table}

\section{An example of soap production}
\label{Sec:app}

We analysis the performance of the different lasso methods by applying to data from a soap processing plant, discussed in \cite{GBM}. This scenario involves examining the output based on the soap mixture components ($x_1$, $x_2$, $x_3$) under specific constraints:
\begin{equation}
	0.2 \leq x_1 \leq 0.8, \quad 0.15 \leq x_2 \leq 0.5, \quad 0.05 \leq x_3 \leq 0.3, \quad x_1 + x_2 + x_3 = 1.
\end{equation}
The process variables of interest include the mixing time ($w_1$) and plodder temperature ($z_1$), and the humidity ($z_2$) with the two latter being harder to control and considered as noise.

Consider the model in \eqref{generalmodel} in its linear version. The matricial formulation is given by,
\begin{equation}
    \underline{x}=\begin{bmatrix}
 x_1 \\
 x_2 \\
 x_3 \\
\end{bmatrix},  \quad w = [w], \quad  \underline{z}=\begin{bmatrix}
 z_1 \\
 z_2 \\
\end{bmatrix},
 \quad   V=\begin{bmatrix}
 w & 0 \\
 0 & w\\
\end{bmatrix} ,
\end{equation}
and
\begin{equation}
\label{parammodel}
 \underline{\alpha}=\begin{bmatrix}
 \alpha_1 \\
 \alpha_2 \\
 \alpha_3 \\ 
\end{bmatrix}, \quad \underline{\delta}=\begin{bmatrix}
 \delta_1 \\
 \delta_2 \\
 \delta_3 \\ 
\end{bmatrix}, \quad \Delta=\begin{bmatrix}
 \gamma_{11} & \gamma_{12} \\
\gamma_{21} & \gamma_{22} \\
  \gamma_{31} &\gamma_{32}  \\
\end{bmatrix}  \text{  and }    H =\begin{bmatrix}
 \eta_{11} &\eta_{12} \\
 \eta_{21} &\eta_{22}\\
 \eta_{31} &\eta_{32}\\ 
\end{bmatrix},
\end{equation}
where,
\begin{equation}
    Y= f(x,w,z)=\underline{x}' \underline{\alpha}+\underline{x}' \underline{\delta} w +\underline{x}' \Delta \underline{z} +\underline{x}' H V \underline{z} + \epsilon
\end{equation}
Table \ref{tab:valores_eje3} shows the estimation for OLS, proposed in \cite{GBM}, and the proposed regularization methods for all the 18 parameters in \eqref{parammodel}. It is worth noting that \cite{GBM} uses OLS to fit the model, and then performs a statistical significance analysis by iteratively eliminating terms with $p$-values greater than 0.05 from the initial model, until all remaining terms are statistically significant. It is worth mentioning that classical lasso excludes more covariables than other methods. Besides, the CAVI Bayesian lasso fits a model with more covariables.

\begin{figure}
    \centering
        \includegraphics[width=1\textwidth]{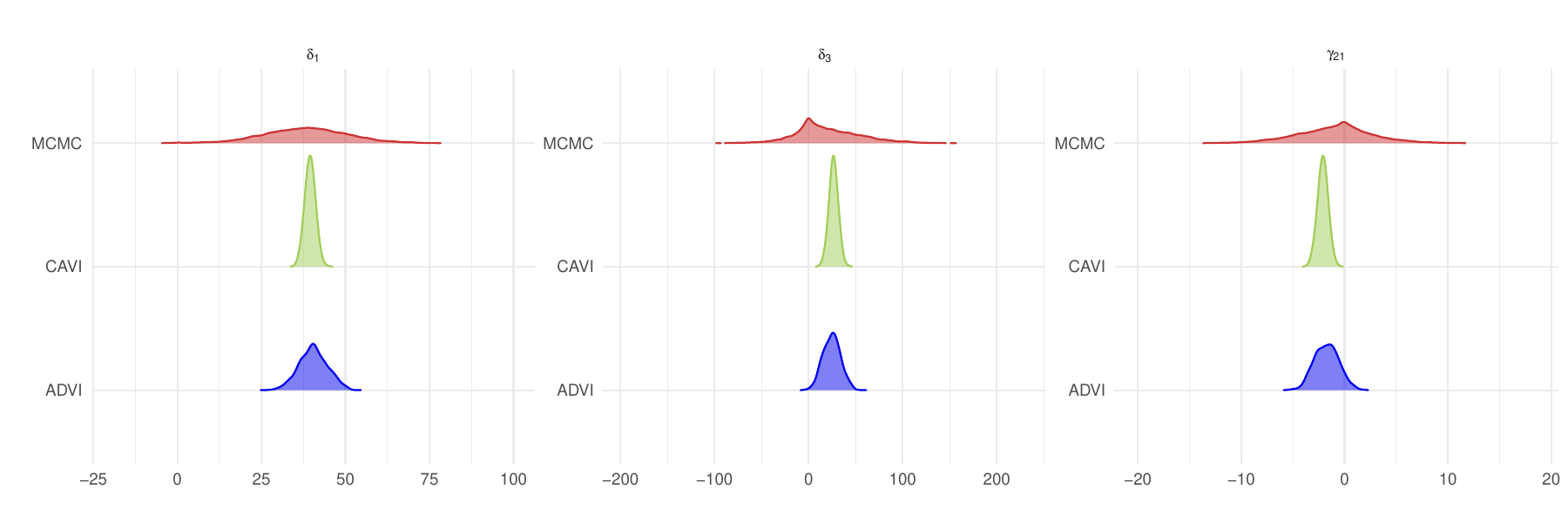}
    \includegraphics[width=1\textwidth]{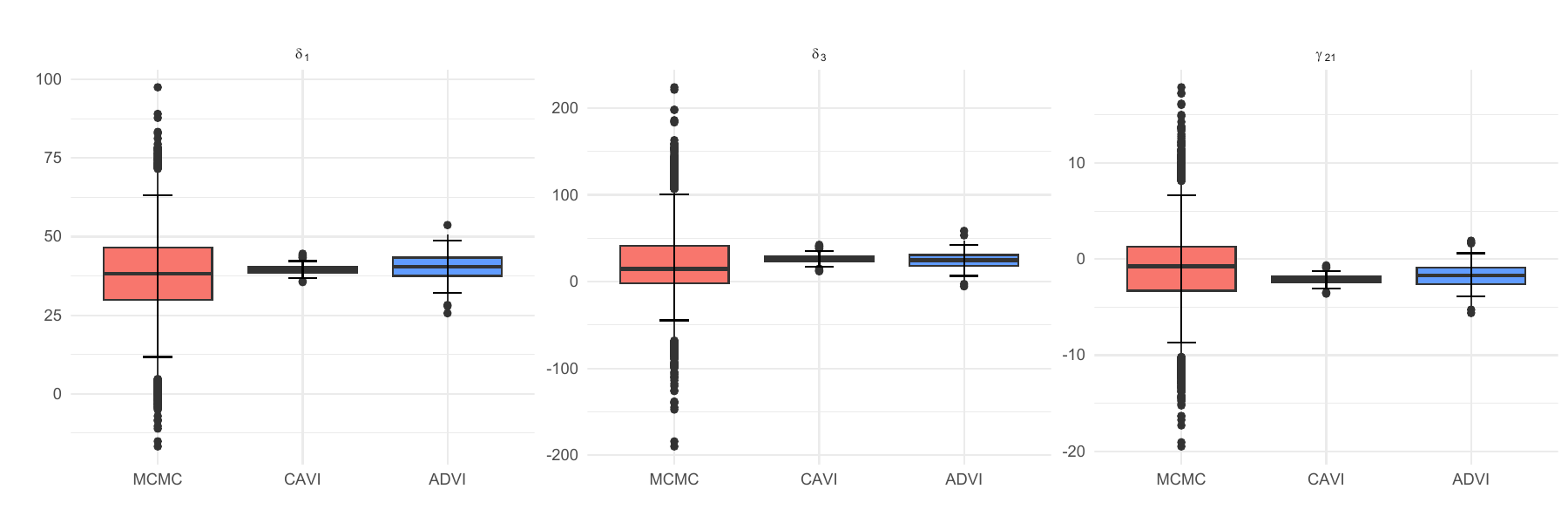}
    \caption{Posterior densities and boxplots of parameters $\delta_1$, $\delta_3$ and $\gamma_{21}$.}
    \label{fig:bloxplotPOS}
\end{figure}

In Figure \ref{fig:bloxplotPOS}, we include the density and bloxplot for posterior distribution of parameter $\delta_1$, which was eliminated by the BL-MCMC using CI criterion. Moreover, the parameters $\delta_3$ and $\gamma_{21}$, which were eliminated solely by the BL-ADVI using the CI criterion. Note that, the posterior densities for such parameters show that the MCMC has more variability, and CAVI is the most homogeneous one, concentrating the probability in a region that does not contain the value zero.

Given the model's 18 parameters and the 40 observations, leave-one-out cross-validation (LOO CV) is particularly advantageous for evaluating our model ( see \cite{Wong}). LOO CV optimizes data utilization and provides a precise, unbiased estimate of the generalization error, essential for reliable performance assessment in this context. Table \ref{tab:LOOCS} presents the LOO CV results for OLS, classical and Bayesian lasso regularization methods. These results demonstrate the comparative effectiveness of each method in minimizing the generalization error, highlighting the robustness of the Bayesian approaches, particularly the BL-CAVI method, which achieved the lowest LOO CV value.

\begin{table}[ht]
\centering
\begin{tabular}{|c|c|c|c|c|c|c|c|c|}
\hline
\multirow{2}{*}{Parameter} & \multirow{2}{*}{OLS} & \multirow{2}{*}{LASSO} & \multicolumn{2}{c|}{BL-MCMC} & \multicolumn{2}{c|}{BL-CAVI} & \multicolumn{2}{c|}{BL-ADVI} \\[0.1cm]
 \cline{4-9}
 &  & &   CI & SN &    CI & SN &    CI & SN \\
\hline
& & & \multicolumn{2}{c|}{} & \multicolumn{2}{c|}{} & \multicolumn{2}{c|}{} \\[-0.3cm]
$\widehat{\alpha}_1$ & 1898.99 &1928.32 & \multicolumn{2}{c|}{1900.10} & \multicolumn{2}{c|}{1899.02} & \multicolumn{2}{c|}{1897.50} \\
$\widehat{\alpha}_2$ & 1626.42 & 1699.97 & \multicolumn{2}{c|}{1624.27} & \multicolumn{2}{c|}{1627.03} & \multicolumn{2}{c|}{1624.75} \\
$\widehat{\alpha}_3$ & 1537.79 & 1431.18 & \multicolumn{2}{c|}{1541.43} & \multicolumn{2}{c|}{1536.33} & \multicolumn{2}{c|}{1535.70} \\
$\widehat{\delta}_1$ & 39.53 & - & \ \ \ - \ \ \ & 38.07 & \multicolumn{2}{c|}{39.52} & \multicolumn{2}{c|}{40.40} \\
$\widehat{\delta}_2$ & 285.90 & 170.58 & \multicolumn{2}{c|}{288.79} & \multicolumn{2}{c|}{285.08} & \multicolumn{2}{c|}{284.81} \\
$\widehat{\delta}_3$ & - & - & \multicolumn{2}{c|}{-} & \multicolumn{2}{c|}{26.57} & - & 24.52 \\
$\widehat{\gamma}_{11}$ & 9.45 & 6.59 & \multicolumn{2}{c|}{9.42} & \multicolumn{2}{c|}{9.45} & \multicolumn{2}{c|}{9.45} \\
$\widehat{\gamma}_{21}$ & - & 2.47 & \multicolumn{2}{c|}{-} & \multicolumn{2}{c|}{-2.08} & \ \ \ - \ \ \ & -1.74 \\
$\widehat{\gamma}_{31}$ & 34.60 & - & \multicolumn{2}{c|}{31.95} & \multicolumn{2}{c|}{34.16} & \multicolumn{2}{c|}{33.99} \\
$\widehat{\gamma}_{12}$ & - & - & \multicolumn{2}{c|}{-} & \multicolumn{2}{c|}{-} & \multicolumn{2}{c|}{-} \\
$\widehat{\gamma}_{22}$ & -20.00 & -16.87 & \multicolumn{2}{c|}{-20.00} & \multicolumn{2}{c|}{-20.00} & \multicolumn{2}{c|}{-20.02} \\
$\widehat{\gamma}_{32}$ & - & - & \multicolumn{2}{c|}{-} & \multicolumn{2}{c|}{-} & \multicolumn{2}{c|}{-} \\
$\widehat{\eta}_{11}$ & 16.84 & 10.15 & \multicolumn{2}{c|}{16.85} & \multicolumn{2}{c|}{16.83} & \multicolumn{2}{c|}{16.69} \\
$\widehat{\eta}_{21}$ & 39.17 & 20.43 & \multicolumn{2}{c|}{37.45} & \multicolumn{2}{c|}{38.91} & \multicolumn{2}{c|}{38.62} \\
$\widehat{\eta}_{31}$ & -21.09 & - & \multicolumn{2}{c|}{-17.64} & \multicolumn{2}{c|}{-20.55} & \multicolumn{2}{c|}{-20.16} \\
$\widehat{\eta}_{12}$ & - & - & \multicolumn{2}{c|}{-} & \multicolumn{2}{c|}{-} & \multicolumn{2}{c|}{-} \\
$\widehat{\eta}_{22}$ & -25.00 & -20.49 & \multicolumn{2}{c|}{-24.99} & \multicolumn{2}{c|}{-24.99} & \multicolumn{2}{c|}{-25.02} \\
$\widehat{\eta}_{32}$ & - & - & \multicolumn{2}{c|}{-} & \multicolumn{2}{c|}{-} & \multicolumn{2}{c|}{-} \\[0.2cm]
\hline
\end{tabular}
\caption{Estimated values for coefficients in \eqref{parammodel}.}
\label{tab:valores_eje3}
\end{table}

\begin{table}[ht]
\begin{center}
\begin{tabular}{| c | c | c | c | c | c | c | c | c | }
\hline
 & \multirow{2}{*}{OLS} & \multirow{2}{*}{LASSO} & \multicolumn{2}{c|}{BL-MCMC} & \multicolumn{2}{c|}{BL-CAVI}& \multicolumn{2}{c|}{BL-ADVI} \\[0.1cm]
 \cline{4-9}
 &  & &   CI & SN & CI & SN & CI & SN \\
\hline
LOO CV &  11.87 & 11.35  & 11.10 &  11.48 & \multicolumn{2}{c|}{ 10.64 }& 11.07  &  11.44 \\
\hline
\end{tabular}
\caption{Values of LOO-CV for OLS and Bayesian lasso versions (MCMC, CAVI and ADVI).}
\label{tab:LOOCS}
\end{center}
\end{table}

\subsection{Optimization of the response surface by the desirability function}

After fitting a combined mixture process-noise variable model, the final goal is to identify levels of the mixture components and controllable variables that simultaneously
yield acceptable mean and variance response values. To address this optimization problem, we use the desirability function approach as outlined in \eqref{di} and \eqref{D}.
Following the methodology described by \cite{GBM}, we initially assume that the noise variables have zero-mean. Under this assumption, we utilize the delta method, applying a first-order Taylor series approximation around the mean of the noise variables. 
Therefore, the expected value and variance of $Y$ are approximated as follows:
\begin{equation}
	\mathbb{E}(Y) \sim \underline{x}'\underline{\alpha} + \underline{x}'\textbf{A}\underline{w},
\end{equation}
and
\begin{equation}
	\mathbb{V}(Y)\sim \left [ \Lambda^{'} \underline{X} + V' \Lambda \underline{X} \right ]^{'} \Sigma_{X} \left [ \Lambda^{'} \underline{X} + V' H \underline{X} \right ].
\end{equation}
These equations allow us to quantify how the proportions of the mixture components and the process conditions directly affect the variability and predictability of the response.
Finally, for OLS and the proposed regularization methods, we show the optimal values in Table \ref{tab:optimosEJ3}. It is evident that all the methods yield identical proportions for the mixture components, yet there are notable differences in the values of the process variable. In terms of the expected value of the response variable, the Bayesian methods utilizing the SN criterion perform better. However, when also considering the variance of the response variable, the CAVI approximation emerges as the superior choice, exhibiting the smallest coefficient of variation.

\begin{table}[ht]
	\begin{center}
		\begin{tabular}{|c |c |c |c |c |c |c |c |c |}  \hline

\multirow{2}{*}{ } & \multirow{2}{*}{OLS} & \multirow{2}{*}{LASSO} & \multicolumn{2}{c|}{BL-MCMC} & \multicolumn{2}{c|}{BL-CAVI}& \multicolumn{2}{c|}{BL-ADVI} \\[0.1cm]
\cline{4-9}
&  & &   CI & SN & CI & SN & CI & SN \\
\hline
& & & \multicolumn{2}{c|}{} & \multicolumn{2}{c|}{} & \multicolumn{2}{c|}{} \\[-0.3cm]
			$x_1$ & 0.45 & 0.45 & \multicolumn{2}{c|}{0.45} &\multicolumn{2}{c|}{0.45} & \multicolumn{2}{c|}{0.45}\\ 
			$x_2$ & 0.50 & 0.50 & \multicolumn{2}{c|}{0.50} & \multicolumn{2}{c|}{0.50} &\multicolumn{2}{c|}{0.50} \\ 
			$x_3$ & 0.05 & 0.05 & \multicolumn{2}{c|}{0.05} &  \multicolumn{2}{c|}{0.05} &  \multicolumn{2}{c|}{0.05} \\  \hline
			$w$ & 0.85 & 0.91 &  \multicolumn{2}{c|}{0.89} &  \multicolumn{2}{c|}{0.90 }& 0.87 & 0.91  \\ \hline  \hline
   & & & & & \multicolumn{2}{c|}{} & & \\[-0.3cm]
    
			$\widehat{\mu_Y}$ & 1881.28 & 1884.35 & 1872.76 & 1888.01 &  \multicolumn{2}{c|}{1901.13} & 1882.75 & 1890.29 \\ 
			$\widehat{\sigma_Y}$ & 15.59 & 15.78 & 15.62 & 15.62&  \multicolumn{2}{c|}{15.50} & 15.66 & 15.61  \\ 
   	$CV$ & 0,00828 & 0,00837 &  0,00834 &0,00827 &  \multicolumn{2}{c|}{0,00815} &0,00831 & 0,008257 \\
        \hline
		\end{tabular}
		\caption{Optimal values for OLS and Bayesian lasso versions (MCMC, CAVI and ADVI). CV = $\widehat{\sigma_Y}/\widehat{\mu_Y}$.}
		\label{tab:optimosEJ3}
	\end{center}
\end{table}





\section{Conclusions}
\label{Sec:con}
We proposed the use of classical and Bayesian lasso regularization for mixture experiments with noise variables. The model formulation was given in \eqref{generalmodel}, where optimization was pursued through the desirability function method \cite{DS}, aiming to simultaneously maximize the mean and minimize the variance of the response variable.

The findings from our study highlight the efficacy of proposed regularization techniques in the context of mixture experiments with noise variables. The comparative analysis, which included ordinary least squares (OLS), lasso, and various Bayesian lasso formulations (CAVI, ADVI, and MCMC), demonstrated the superior performance of the CAVI algorithm. CAVI consistently outperformed other methods in both simulation studies and real data applications, particularly in terms of variable selection accuracy and response surface optimization. In a practical application involving a soap processing plant, CAVI not only provided precise parameter estimates but also optimized the response, achieving higher expected values and lower variance. Bayesian lasso variants, especially CAVI and ADVI, proved advantageous over classical lasso in robustness and flexibility of parameter estimation. While MCMC-based methods were reliable, they faced challenges related to convergence and computational efficiency in high-dimensional spaces. Variational inference methods (CAVI and ADVI) offered efficient approximations to posterior distributions, significantly reducing computational time compared to MCMC. The application of Bayesian regularization techniques, particularly CAVI, enhances model selection, parameter estimation, and response optimization in complex systems influenced by both mixture components and process variables.

\section*{Acknowledgements} MGN was partially supported by Fondecyt Iniciación 11200500.


\end{document}